\documentclass[12pt]{article}
\usepackage{latexsym}
\usepackage{graphicx}
\thicklines                         % thick straight lines for pictures (LaTeX)
\long \def \blockcomment #1\endcomment{}
\begin{document}           % End of preamble and beginning of text.
\baselineskip=0.33333in

%\begin{quote} \raggedleft TAUP 2871-2007
%\end{quote}
%\title{A Sample Document}  % Declares the document's title.
%\author{Leslie Lamport}    % Declares the author's name.
%\date{December 12, 1984}   % Deleting this command produces today's date.
\vglue 0.5in
%\maketitle                 % Produces the title.
\begin{center}{\bf On the Significance of the Upcoming \\
LHC {\em pp} Cross Section Data}
\end{center}
\begin{center}E. Comay$^*$
\end{center}

\begin{center}
Charactell Ltd. \\
PO Box 39019 \\
Tel-Aviv, 61390, Israel
\end{center}
\vglue 0.5in
% \vglue 0.5in
\vglue 0.2in
\noindent
Abstract:

The relevance of the Regular Charge-Monopole Theory to the
proton structure is described. The discussion relies on classical
electrodynamics and its associated quantum mechanics. Few
experimental data are used as a clue to the specific structure
of baryons. This basis provides an explanation for
the shape of the graph of the
pre-LHC proton-proton cross section data.
These data also enable a description of
the significance of the expected LHC cross section measurements which
will be known soon. Problematic QCD issues are pointed out.

\newpage
\noindent
{\bf 1. Introduction}
\vglue 0.33333in

Scattering experiments are used as a primary tool for investigating
the structure of physical objects. These experiments can be divided
into several classes, depending on the kind of colliding particles.
The energy involved in
scattering experiments has increased dramatically
during the previous century since the celebrated Rutherford experiment
was carried out (1909). Now, the meaningful
value of scattering energy is the quantity measured in the
rest frame of the projectile-target center of energy.
Therefore, devices that use colliding beams enable measurements of
very high energy processes. The new Large Hadron Collider
(LHC) facility
at CERN, which is designed to produce 14 TeV proton-proton ($pp$)
collisions, will make a great leap forward.

This work examines the presently available $pp$ elastic and total
cross section data (denoted by ECS and TCS, respectively)
and discusses the meaning of two possible alternatives for
the LHC $pp$ ECS values which will be known soon.
The discussion relies
on the Regular Charge-Monopole Theory (RCMT) [1,2] and on its
relevance to strong interactions [3,4].

Section 2 contains a continuation of the discussion presented in [4].
It explains the meaning of two possible LHC results of the
$pp$ ECS. Inherent QCD difficulties to provide an explanation for the data
are discussed in section 3. The last section contains concluding remarks.

\vglue 0.66666in
\noindent
{\bf 2. The Proton-Proton Elastic Cross Section}
\vglue 0.33333in

The discussion carried out below is a continuation of [4].
Here it aims to examine possible LHC's ECS results and their
implications for the proton structure.
Thus, for the reader's convenience, the relevant points of [4] are
presented briefly in the following lines.

RCMT is the theoretical basis of the discussion and strong interactions
are regarded as interactions between magnetic monopoles which obey
the laws derived from RCMT.
Two important results of RCMT are described here:

\begin{itemize}
\item[{1.}]
Charges do not interact with bound fields of monopoles and
monopoles do not interact with bound fields of charges. Charges interact
with all fields of charges and with radiation fields emitted from
monopoles. Monopoles interact with all fields of monopoles and with
radiation fields emitted from charges.

\item[{2.}]
The unit of the elementary magnetic charge $g$ is a
free parameter. However, hadronic data indicate that
this unit is much larger than that of the electric charge:
$g^2 \gg e^2 \simeq 1/137$. (Probably $g^2 \simeq 1$.)
\end{itemize}

\begin{figure}[t]
  \centering
    \includegraphics[width=8cm]{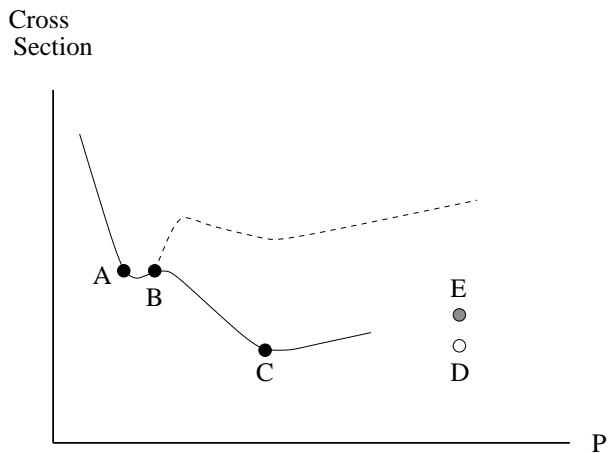}
  \caption{\em
A qualitative description of
the pre-LHC proton-proton cross section versus the
laboratory momentum P. Axes are drawn in a logarithmic scale.
The solid line denotes
elastic cross section and the broken line denotes total cross section.
(The accurate figure can be found in [5]). Points A-E help the
discussion (see text).
}
\end{figure}

The application of RCMT to strong interactions regards quarks
as spin-1/2 Dirac particles that carry a magnetic monopole unit.
A proton has three valence quarks {\em and} a core that carries
three monopole units of the opposite sign. Thus, a proton is a magnetic
monopole analogue of a nonionized atom. By virtue of
the first RCMT result, one understands why electrons
(namely, pure charges) do not participate in strong interactions
whereas photons do that [6]. Referring to the pre-LHC data,
it is shown in [4] that,
beside the three valence quarks, a proton has a core that contains
inner closed shells of quarks.

Applying the correspondence between a nonionized atom
and a proton, one infers the validity of screening effects and of
an analogue of the Franck-Hertz effect that takes place for the
proton's quarks. Thus,
quarks of closed shells of the proton's core behave like inert
objects for cases where the projectile's energy is smaller than
the appropriate threshold.

The pre-LHC $pp$ scattering data is depicted in fig. 1. Let $ep$
denote both electron-proton and positron-proton interaction.
Comparing the $ep$ scattering data with those of $pp$, one finds
a dramatic difference between both the ECS and the TCS characteristics
of these experiments.
Thus, the deep inelastic and the Rosenbluth $ep$ formulas respectively
show that TCS {\em decreases} together with an increase
of the collision energy and that at the high energy region, ECS
decreases even faster and
takes a negligible part of the entire TCS events (see [7], p. 266).
The $pp$ data of fig. 1 show a completely different picture.
Indeed, for high
energy, both the TCS and the ECS $pp$ graphs {\em go up} with collision
energy and ECS takes about $15\%$ of the total events.

The last property proves that a proton contains a quite solid
component that can take the heavy blow of a high energy
$pp$ collision and
leave each of the two colliding protons intact. Valence quarks certainly
cannot do this, because in the case of a
high energy $ep$ scattering, an electron
collides with a valence quark. Now, in this case, deep inelastic
scattering dominates and elastic events are very rare.
The fact that the quite solid
component is undetected in an $ep$ scattering experiment, proves that it is
a spinless electrically neutral component. This outcome provides
a very strong support for the RCMT interpretation of hadrons,
where baryons have a core [3,4].

The foregoing points enable one to interprete the
shape of the $pp$ ECS graph of fig. 1.
Thus, for energies smaller than that of point A of the figure, the
wave length is long and effects of large distance between
the colliding protons dominate the
process. Here the ordinary Coulomb potential, $1/r$, holds and
the associated $1/p^2$ decrease of the graph is in accordance
with the Rutherford and Mott formulas (see [7], p. 192)
\begin{equation}
(\frac {d\sigma }{d\Omega })_{Mott} =
\frac {\alpha ^2 \cos ^2 (\theta /2)}
{4p^2 \sin ^4 (\theta /2)[1 + (2p/M) \sin^2 (\theta /2)]},
\label{eq:MOTT}
\end{equation}
At the region of points A-B,
the rapidly varying nuclear force makes the undulating shape of
the graph. Results of screening effects of the valence quarks
are seen for momentum values belonging to the region of points B-C.
Indeed, a correspondence holds for electrons in an atom and
quarks (that carry a monopole unit) in a proton.
Hence, for a core-core interaction, the
screening associated with the valence quarks
weakens as the distance from the proton's center
becomes smaller. It means that
the strength of the core's monopole potential arises faster than the
Coulomb $1/r$ formula. For this reason, the decreasing slope of
the graph between points B-C is smaller than that which is seen
on the left hand side of point A.

The ECS graph stops decreasing and
begins to increase on the right hand side of point C.
This change of the graph's slope indicates that
for this energy
a new effect shows up. Indeed, assume that the proton consists of just valence
quarks and an elementary pointlike core which is
charged with three monopole units of the opposite sign. Then,
as the energy increases and the wave length decreases, the contribution
of the inner proton region becomes more significant. Now, at inner regions,
the valence quarks'
screening effect fades away and the potential tends to the Coulomb
formula $1/r$. Hence, in this case, the steepness of the decreasing
graph between points B-C
{\em should increase} near point C and tend to the
Coulomb-like steepness of
the graph on the left hand side of point A. The data negate this
expectation. Thus, the increase of the graph on the right hand
side of point C
indicates the existence of inner closed shells of quarks
at the proton. It is concluded that at these shells, a new screening
effect becomes effective.

It is interesting to note that at the same momentum region also the TCS
graph begins to increase and that on the right hand side of point C,
the vertical distance between the two graphs is uniform. The logarithmic
scale of the figure proves that, at this region,
the ratio ECS to TCS practically does
not change. The additional TCS events are related to an analogue
of the Franck-Hertz effect. Here a quark of the closed shells is struck
out of its shell.
This effect corresponds to the $ep$ deep inelastic process and it
is likely to produce an inelastic event.

The main problem to be discussed here is {\em the specific structure of the
proton's closed shells of quarks.} One may expect that the situation
takes the simplest case and that the core's closed shells consist of
just two $u$ quarks and two $d$ quarks that occupy an S shell. The other
extreme is the case where the proton is analogous to a very heavy atom
and the proton's core contains many closed shells of quarks. Thus, the
energy of the higher group of the core's shells
takes quite similar value and their radial wave functions
partially overlap. (Below,
finding the actual structure of the proton's core is called Problem $A$.)
The presently known $pp$ ECS data which
is depicted in fig. 1 is used for describing the relevance of the LHC
future data to Problem $A$.

The rise of the $pp$ ECS graph on the right hand side of point C
is related to a screening effect of the proton's inner closed shells. Now,
if the simplest case which is described above holds then,
for higher energies, this effect should diminish and
the graph is expected to stop rising and pass near the open circle
of fig. 1, which is marked
by the letter D. On the other hand, if the proton's core contains
several closed shells having a similar energy and
a similar radial distribution,
then before the screening contribution of
the uppermost closed shell fades away another shell is expected to enter
the dynamics. In this case, the graph is expected to continue rising
up to the full LHC energy and
pass near the gray circle of fig. 1, which is marked by the letter E [8].

The foregoing discussion shows one example explaining how the LHC
data will improve our understanding of the proton's structure.

\vglue 0.66666in
\noindent
{\bf 3. Inherent QCD Difficulties}
\vglue 0.33333in

Claims stating that QCD is unable to provide an explanation for
the $pp$ cross section data have been published in the last decade [9].
Few specific reasons justifying these claims are listed below.
The examples rely on QCD's
main property where baryons consist of three valence quarks, gluons
and possible pairs of quark-antiquark.

\begin{itemize}

\item Deep inelastic $ep$ scattering proves that for a very high energy,
elastic events are very rare (see [7], p. 266).
It means that an inelastic event is found
for nearly every case where a
quark is struck violently by an electron.
On the other hand, fig. 1 proves that for high energy,
elastic $pp$ events take about
$15\%$ of the total events.
Therefore, one wonders what is the proton's component that takes the
heavy blow of a high energy $pp$ collision and is able to
leave the two colliding protons intact?
Moreover, why this component is not observed in the corresponding $ep$
scattering?

\item A QCD property called Asymptotic Freedom (see [10], p. 397)
states that the
interaction strength tends to zero at a very small vicinity of a QCD
particle. Thus, at this region, a QCD interaction is certainly much
weaker than the corresponding Coulomb-like interaction. Now,
the general expression for the elastic
scattering amplitude is (see [7], p. 186)
\begin{equation}
M_{if} = \int \psi ^*_f V\psi _i d^3x,
\label{eq:MIF}
\end{equation}
where $V$ represents the interaction.
Evidently, for very high energy, the contribution of a very short
distance between the colliding particles dominates the process.
Therefore, if asymptotic freedom holds then the $pp$
ECS line is expected to show a {\em steeper
decrease} than that of the Coulomb interaction, which is seen on
the left hand side of point A of fig. 1. The data represented in
fig. 1 proves that for an energy which is
greater than that of point C of fig. 1, the
$pp$ ECS line {\em increases}. Hence, the
data completely contradict this QCD property.

\item A general argument. At point C of fig. 1, the ECS
graph changes its inclination. Here it stops decreasing and
begins to increase. This effect proves that for this energy value,
{\em something new shows up in the proton.} Now, QCD states that quarks
and gluons are elementary particles that move quite freely inside
the proton's volume. Therefore, one wonders how can QCD explain why a new
effect shows up for this energy?

\end{itemize}

Each of these specific points illustrates the general statement of [9],
concerning QCD's failure to describe the high energy $pp$
cross section data.

\vglue 0.66666in
\noindent
{\bf 4. Concluding Remarks}
\vglue 0.33333in

The following lines describe the logical structure of this work and thereby
help the reader to evaluate its significance.

A construction of a physical theory must assume the
validity of some properties of the physical world.
For example, one can hardly imagine how can a person
construct the Minkowski space with
{\em three} spatial dimensions,
if he is not allowed to use experimental data. Referring to
the validity of a physical theory, it is well known that
unlike a mathematical theory which is evaluated just
by pure logics, a physical
theory must also be consistent with well established experimental data
that belong to its domain of validity.
The Occam's razor principle examines another aspect of a theory
and prefers a theory that
relies on a minimal number of assumptions. Thus, the Occam's razor
can be regarded as a "soft" acceptability criterion for a theory.

Following these principles, the assumptions used for
the construction of RCMT and of its application to strong interactions
are described below. The first point has a theoretical
character and the rest rely on experimental results that serve
as a clue for understanding the specific structure of baryons.

\begin{itemize}

\item A classical regular charge-monopole theory is constructed on the
basis of duality relations between ordinary Maxwellian theory of
charges together with their fields and a monopole system together
with its associated
fields [2]. (In [1], it is also required that the theory be derived
from a regular Lagrangian density.) Like ordinary electrodynamics,
this theory is derived from the variational principle
where regular expressions are used. Hence, the
route to quantum mechanics is straightforward.

\item
In RCMT, the value of the elementary monopole unit $g$ is a free parameter.
Like the case of the electric charge, it is assumed that
$g$ is quantized. It is also assumed that its elementary value
$g^2 \gg e^2 \simeq 1/137$. (Probably, $g^2 \simeq 1$.)

\item It is assumed that strong interactions are interactions between
monopoles. The following points describe the specific systems that carry
monopoles.

\item It is assumed that quarks are spin 1/2 Dirac
particles that carry a unit of magnetic monopole [11].

\item It is assumed that baryons contain {\em three} valence quarks. It
follows that baryons must have a core that carries three monopole units
of the opposite sign.

\item It is assumed that the baryonic core contains
closed shells of quarks.

\end{itemize}

The discussion carried out in [4] and in
section 2 of this work explains how RCMT can be used
for providing a qualitative interpretation of the shape
of the graph that describes the elastic $pp$ scattering data.
In particular, an explanation is provided for the relation between
the pre-LHC {\em pp} elastic cross section data and the existence of closed
shells of quarks at the baryonic core. It is also explained how the
upcoming LHC data will enrich our understanding of the
structure of baryonic closed shells of quarks by providing information
on whether there are just two active closed shells of $u$ and $d$
quarks or there are many shells having a quite similar energy value
and radial distribution.

QCD's inherent difficulties to provide an explanation for the
high energy pre-LHC $pp$ scattering data are discussed in the third section.
Screening effects of proton's quarks are used in the Regular Charge-monopole
Theory's interpretation of the elastic cross section $pp$ scattering. It
is interesting to note that this kind of screening also provides
an automatic explanation for the first EMC effect [12]. This effect
compares the quarks' Fermi motion in deuteron and iron (as well as other
heavy nuclei). The data show that the Fermi motion is smaller in
hevier nuclei. This experimental data and
the Heisenberg uncertainty relations prove that the quarks'
self-volume increases in heavier nuclei. In spite of the
quite long time elapsed, QCD supporters have not yet provided an
adequate explanation for the first EMC effect [13].

\newpage
References:
\begin{itemize}
\item[{*}] Email: elicomay@post.tau.ac.il  \\
\hspace{0.5cm}
           Internet site: http://www.tau.ac.il/$\sim $elicomay

\item[{[1]}] E. Comay {\em Nuovo Cimento B} {\bf 80}, 159 (1984).
\item[{[2]}] E. Comay {\em Nuovo Cimento B} {\bf 110}, 1347 (1995).
\item[{[3]}] E. Comay
{\em A Regular Theory of Magnetic Monopoles and Its Implications} in
{\em Has the Last Word Been Said on Classical Electrodynamics?}
ed. A. Chubykalo, V. Onoochin, A. Espinoza and R. Smirnov-Rueda
(Rinton Press, Paramus, NJ, 2004).
\item[{[4]}] E. Comay {\em Apeiron} {\bf 16}, 1 (2009).
\item[{[5]}] C. Amsler et al., Phys. Lett. {\bf B667}, 1 (2008).
(See p. 364).
\item[{[6]}] T. H. Bauer, R. D. Spital, D. R. Yennie
and F. M. Pipkin, Rev. Mod. Phys. {\bf 50}, 261 (1978).
\item[{[7]}] D. H. Perkins, {\em Introduction to High Energy Physics}
(Addison-Wesley, Menlo Park CA, 1987).
\item[{[8]}] This possibility has been overlooked in [4].
\item[{[9]}] A. A. Arkhipov,
\\
\nolinebreak
{http://arxiv.org/PS\_cache/hep-ph/pdf/9911/9911533v2.pdf}
\item[{[10]}] H. Frauenfelder and E. M. Henley, {\em Subatomic Physics},
(Prentice Hall, Englewood Cliffs, 1991). pp. 296-304.
\item[{[11]}] As a matter of fact, it can be {\em proved} that an
elementary massive quantum mechanical particle is a spin-1/2
Dirac particle. See: E. Comay, Progress in Physics, {\bf 4}, 91 (Oct. 2009).
\item[{[12]}] J. J. Aubert et al. (EMC), Phys. Lett., {\bf 123B}, 275 (1983).
\item[{[13]}] J. Arrington et al., J. Phys. Conference Series,
{\bf 69}, 012024 (2007).

\end{itemize}

\end{document}